\documentclass[fleqn]{elsart}
\usepackage{amsmath}
\usepackage{graphicx}
\usepackage{caption2}
\begin{document}

\begin{frontmatter}
\title{Hardy-type experiment for the maximally entangled state:
Illustrating the problem of subensemble postselection\thanksref{paper}}

\author{Jos\'{e} L. Cereceda}
\address{C/Alto del Le\'on 8, 4A, 28038 Madrid, Spain}
\thanks[paper]{This paper has been originally published in: J.L. Cereceda, Phys. Lett. A 263, 232-244 (1999).}

\vspace{.5cm}
\begin{abstract}
By selecting a certain subensemble of joint detection events in a
two-particle interferometer arrangement, a formal nonlocality
contradiction of the Hardy type is derived for an ensemble of particle
pairs configured in the maximally entangled state. It is argued,
however, that the class of experiments exhibiting this kind of
contradiction does not rule out the assumption of local realism.

\vspace{.5cm}
\begin{flushleft}
nonlocality theorem; Bell's inequality; Interaction-free measurement
\end{flushleft}

\end{abstract}
\end{frontmatter}

\vspace{-.25cm}

\section{Introduction}

A remarkable feature of Hardy's nonlocality theorem
\cite{Hardy93,Goldstein94} is that it applies to any pure entangled
states of a bipartite two-level system, except, curiously, those
states which are maximally entangled such as the singlet state of two
spin-half particles. As Hardy says \cite{Hardy93}, ``The reason for
this is that the proof relies on a certain lack of symmetry that is
not available in the case of a maximally entangled state.'' Indeed,
for this class of states, every single-particle observable is
perfectly, symmetrically correlated with some other observable
associated with the other particle. As a result, the set of Hardy
equations (\ref{H1})-(\ref{H4}) (see below) upon which the nonlocality
contradiction is constructed cannot be satisfied for the maximally
entangled case \cite{CB,Cereceda98}. An attempt to extend Hardy's
theorem to cover maximally entangled state was made by Wu et al.\ in
Ref.\ \cite{Wu et al.} where the authors, using a quantum-optical
setting, demonstrate local-realism violations for a maximally
entangled state of two particles without using inequalities. It was
later pointed out \cite{Cereceda97a}, however, that the nonlocality
argument in \cite{Wu et al.} requires a minimum total of six
dimensions in Hilbert space, and so there is no contradiction with the
fact that no Hardy-type nonlocality argument can be constructed for
the maximally entangled state if the observables to be measured on
each particle are truly dichotomic. The extra dimensions needed to
properly describe the experiment in Ref.\ \cite{Wu et al.} arise due
to the use of three independent detectors for each of the
particles. Moreover, the probabilistic nonlocality contradiction
derived in \cite{Wu et al.} is conditioned on the fact that the
statistical analysis involved is restricted to a particular
subensemble of joint detection events. The authors define this
subensemble by saying that \cite{Wu et al.}, ``We shall only be
interested in those runs of the experiment for $K = L = 0$, which
means that particle 1 does not go to end $k$, while at the same time
particle 2 does not go to end $l$.'' Although there are situations in
which subensemble selection may be a legitimate means to observe
local-realism violations \cite{Yurke-Stoler,Popescu+Gisin,Peres96}, we
will see that this is not the case for the class of experiments
considered in this Letter. Specifically, by using an arrangement for
two-particle interferometry, we will show that, (a) if each particle
is subjected to a single ideal (von Neumann-type) measurement (chosen
at random between two such possible measurements), then it is
necessary to perform subensemble postselection (or, equivalently, to
reject some `undesirable' subset of measurement data) if we want to
obtain a Hardy-type nonlocality contradiction for the maximally
entangled state; and (b) this procedure to get local-realism
violations does not constitute a valid method to rule out all possible
local hidden variable models since, for the class of experiments
discussed, no Bell-type inequality is violated if the whole ensemble
of measurement data is included in the statistical analysis. The
conclusion to be drawn from these two statements is that the class of
experiments adhering to \mbox{Wu et al.'s} approach does not provide a
true test of quantum mechanics versus local realism, not even in the
case of ideal behaviour of the experimental hardware.

The Letter is organized as follows. In Section 2 we consider a
two-particle interferometer arrangement where a source emits pairs of
particles, 1 and 2, in some quantum-mechanical superposition
state. The outgoing particle 1 (2) is monitored by ideal detectors
$L_1$ and $U_1$ ($L_2$ and $U_2$) so that, for this arrangement, each
particle is subjected to a binary choice between the detection in
either $L_i$ or $U_i$ (where $i=1$ (2) for particle 1 (2)). We will
show that, under this dichotomic choice, no Hardy-type contradiction
can be obtained when the experiment is performed on an ensemble of
particle pairs prepared in the maximally entangled state (\ref{ME})
(see below). In Section 3, the `standard' interferometric arrangement
used in Section 2 is modified so that a partially absorbing material
is placed in one of the routes available to one of the particles (say,
particle 2) inside the interferometer. Naturally, the absorber is a
kind of detector (call it, say, $A_2$) which detects some of the
particles, namely those which are absorbed, while the rest pass
through. Therefore, particle 2 is no longer subjected to a binary
choice since it can be detected in either $L_2$, $U_2$, or $A_2$. The
measurement data we are interested in will now consist of the
subensemble of registration events for which {\em both\/} particles in
a pair end at the corresponding detector $L_i$ or $U_i$, while the
remaining two-particle coincidence detections (namely, those for which
particle 1 ends in either $L_1$ or $U_1$ while particle 2 gets
absorbed before reaching $L_2$ or $U_2$) are discarded. On the other
hand, as shown in Refs.\ \cite{Mermin,Hardy94}, Hardy's nonlocality
argument can be cast in the form of an inequality which is just a
particular case of the Clauser-Horne (CH) inequality
\cite{C-H,C-S}. As we shall see, this inequality is violated for the
above selected subensemble of joint detection events. The amount of
this violation is found to be as large as $\frac{1}{2} \leq
0$. However, since this approach does involve a postselection
procedure, then, it can be justifiably claimed \cite{PHZ} that one
runs directly into the so-called {\em subensemble postselection
problem\/} \cite{C-H,C-S,PHZ,Santos,Caro-G,Peres97} which, in our
case, essentially means that the above local-realism violation would
not be truly significant, as a Bell inequality could always be
violated (even by purely classical correlations \cite{Peres97}) if one
restricts the analysis to a suitable subensemble of the original
ensemble of particle pairs. In Section 4 it is shown that, in fact, no
CH inequality is violated if the entire pattern of localization
correlations is analysed. Finally, in Section 5, we show how our
interferometer set-up can be used to perform an interaction-free
measurement of the presence of the absorber. Conclusions are presented
in Section 6.

\section{Failure of Hardy's proof for the maximally entangled state in the standard interferometer set-up}

In what follows we specialize in photons, although any other suitable
interfering particle could equally be considered. For the present
purpose, we consider a two-photon interferometer arrangement of the
kind first discussed by Horne and Zeilinger \cite{Horne-Zeilinger}
(see Fig.\ 1).
\begin{figure}[t]
\centering
\includegraphics[width=4.5in]{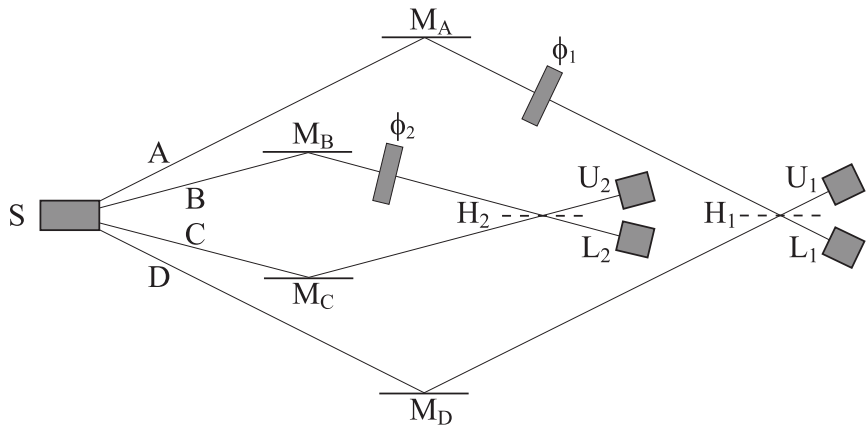}
\renewcommand{\figurename}{Fig.}
\renewcommand{\captionlabeldelim}{.~}
\caption{\footnotesize An arrangement for two-particle interferometry
of the kind proposed in Ref.\ \cite{Horne-Zeilinger}. All four
detectors $L_i$ and $U_i$ ($i=1,2$) are assumed to be 100\%
efficient. Eventually, we will consider a modified version of this
arrangement in which photon 2 has a nonvanishing probability of being
absorbed by the phase shifter placed in path $B$.}
\vskip.5cm
\end{figure}
A parametric down-conversion source is arranged to emit an ensemble of
photon pairs along the beams $A$, $B$, $C$, and $D$, in such a way
that the two photons 1 and 2 in each pair emerge coherently downstream
from the source in the state
\enlargethispage*{1mm}  
\begin{equation}
\left| \psi \right\rangle \,=\,\frac{1}{\sqrt{2}}\left( \left| A\right\rangle
_{1}\left| C\right\rangle _{2}+\left| D\right\rangle _{1}\left| B\right\rangle _{2}\right) ,
\label{ME}
\end{equation}    
where ket ${|A \rangle}_1$ designates photon 1 in beam $A$, etc. Beams
$A$ and $D$ ($B$ and $C$) are totally reflected by respective mirrors
$M_A$ and $M_D$ ($M_B$ and $M_C$), and subsequently recombined at the
(nonpolarising) beam splitter $H_1$ ($H_2$) from which photon 1 (2)
proceeds either to detector $L_1$ or $U_1$ ($L_2$ or $U_2$). Figure~1
also shows two adjustable phase shifters $\phi_1$ and $\phi_2$ placed
into beams $A$ and $B$, respectively. For the initial state
(\ref{ME}), and for symmetric, 50-50 beam splitters and lossless
optical elements, the probabilities of joint detection of the two
photons by the various detectors are, in an obvious notation,
\begin{align}
P(L_{1},\phi_1;L_{2},\phi_2) &= P(U_{1},\phi_1;U_{2},\phi_2)=%
\tfrac{1}{4} \left[ 1+\cos (\phi _{1}-\phi _{2})\right] , \tag{2a}  \label{P1}  \\
P(L_{1},\phi_1;U_{2},\phi_2) &= P(U_{1},\phi_1;L_{2},\phi_2)=%
\tfrac{1}{4} \left[ 1-\cos (\phi _{1}-\phi _{2})\right] .  \tag{2b}  \label{P2}
\end{align}

It is easy to show that, in these cirscumstances, no Hardy-type
contradiction can be obtained for the maximally entangled state
(\ref{ME}). For the interferometric arrangement we are considering,
Hardy's argument for nonlocality involves four alternative
experimental configurations each of which being characterized by a
particular setting of the phase shifters. Specifically, these four
configurations are defined by the respective settings:
$(\phi_1,\phi_2)$, $(\phi_1,\phi^{\prime}_2)$,
$(\phi^{\prime}_1,\phi_2)$, and $(\phi^{\prime}_1,\phi^{\prime}_2)$,
where the first (second) entry inside each bracket is the value of the
phase shifter acting on beam $A$ ($B$). In terms of joint detection
probabilities, the state of the emerging down-converted pair of
correlated photons will show Hardy-type nonlocality contradiction if
the following four conditions are simultaneously fulfilled for such a
state
\enlargethispage*{.2cm}
\begin{align}
P(L_{1},\phi _{1};L_{2},\phi _{2}) & =  0\, ,  \tag{3a}  \label{H1}  \\
P(U_{1},\phi _{1};U_{2},\phi _{2}^{\prime }) & =  0\, , \tag{3b}  \label{H2}  \\
P(U_{1},\phi _{1}^{\prime };U_{2},\phi _{2}) & =  0\, , \tag{3c}  \label{H3}  \\
P(U_{1},\phi _{1}^{\prime };U_{2},\phi _{2}^{\prime }) & > 0\, .  \tag{3d}  \label{H4}
\setcounter{equation}{3}  
\end{align}
Using predictions (\ref{H1})-(\ref{H4}) one can construct an argument
against the notion of local realism. In short, the argument is like
this \cite{Hardy93}: Suppose that, for a particular run of the
experiment, we get photon 1 at $U_1$ and photon 2 at $U_2$ when the
interferometer set-up is operated in the configuration
$(\phi^{\prime}_1, \phi^{\prime}_2)$. From Eq.\ (\ref{H4}), there is a
nonzero probability for this joint detection event to occur. (We
further assume that the event corresponding to detection of photon 1
in $L_1$ or $U_1$ is spacelike separated from that corresponding to
detection of photon 2 in $L_2$ or $U_2$.) Then, by invoking local
realism, and taking into account the prediction (\ref{H3}), we can
deduce that, had the phase shifter in path $B$ been set to $\phi_2$,
photon 2 would have been detected in $L_2$. Likewise, from prediction
(\ref{H2}), and according to local realism, we can assert that photon
1 would have ended in $L_1$, had the phase shifter in path $A$ been
set to $\phi_1$. Therefore, by combining the assumption of local
realism with the quantum predictions (\ref{H4}), (\ref{H3}), and
(\ref{H2}), we are led to conclude that there is a nonzero probability
that {\em both\/} photons in a pair end in the corresponding
$L$-detector when the configuration is set to $(\phi_1,\phi_2)$. The
remaining quantum prediction in Eq.\ (\ref{H1}), however, tells us
that no joint detection of photons 1 and 2 in $L_1$ and $L_2$ will
take place when the interferometer set-up is arranged to operate in
the configuration $(\phi_1,\phi_2)$. Hence a contradiction between
quantum mechanics and local realism arises without using inequalities.

Now, if conditions (\ref{H1}), (\ref{H2}), and (\ref{H3}) are to be
fulfilled for the state (\ref{ME}), and assuming that the joint
detection probabilities accounting for our experiment are those given
by Eqs.\ (\ref{P1}) and (\ref{P2}), it is necessary that (see Eq.\
(\ref{P1})) $\phi_1 - \phi_2 = n_1\pi$, $\phi_1 - \phi^{\prime}_2 =
n_2\pi$, and $\phi^{\prime}_1 - \phi_2 = n_3\pi$, with $n_1, n_2, n_3
= \pm1, \pm3,\ldots$~. From these three equalities we obtain
immediately that $\phi^{\prime}_1 - \phi^{\prime}_2 = \linebreak
(n_2+n_3-n_1)\pi$. As the number $n_2+n_3-n_1$ is an odd integer, we
have $\cos(\phi^{\prime}_1 - \phi^{\prime}_2) = -1$, and then
$P(U_{1},\phi _{1}^{\prime };U_{2},\phi _{2}^{\prime }) = 0$. So, the
fulfillment of all three conditions (\ref{H1}), (\ref{H2}), and
(\ref{H3}), precludes the fulfillment of the condition in Eq.\
(\ref{H4}), and thus no Hardy-type contradiction will be obtained for
the maximally entangled state if each particle is subjected to a
binary choice between detection in either $L_i$ or $U_i$. This
impossibility can equivalently be established by stating that the
fulfillment of conditions (\ref{H1})-(\ref{H3}) by the state
(\ref{ME}) implies that $P(L_1,L_2)=P(U_1,U_2) = 0$ and
$P(L_1,U_2)=P(U_1,L_2)= 1/2$ for {\em any\/} one of the four
configurations $(\phi_1,\phi_2)$, $(\phi_1,\phi^{\prime}_2)$,
$(\phi^{\prime}_1,\phi_2)$, and
$(\phi^{\prime}_1,\phi^{\prime}_2)$. Clearly this means that, for such
configurations, whenever photon 1 is detected at $L_1$ ($U_1$) then
with certainty photon 2 will be detected at $U_2$ ($L_2$), and vice
versa. So, if one assigns a value `$-1$' (`$+1$') to a count in either
$L_1$ or $L_2$ ($U_1$ or $U_2$), then the expectation value of the
product of the two outcomes over a large number of counts will be
$E(\phi_1,\phi_2)=E(\phi_1,\phi^{\prime}_2)=E(\phi^{\prime}_1,\phi_2)=E(\phi^{\prime}_1,\phi^{\prime}_2)=-1$. These
four perfect correlations, however, saturate the Bell-CHSH inequality
\cite{C-S,CHSH,Bell}
\begin{equation}
\left| E(\phi_1,\phi_2) + E(\phi_1,\phi^{\prime}_2) + E(\phi^{\prime}_1,\phi_2) - E(\phi^{\prime}_1,\phi^{\prime}_2) \right| \leq 2.  
\end{equation}
So it is concluded, once again, that, for the interferometric arrangement under consideration, the maximally entangled state (\ref{ME}) cannot be used to exhibit Hardy-type nonlocality.

\section{Hardy-like proof for the maximally entangled state in the modified interferometer set-up}

Now the question arises about how to modify the `standard'
interferometer set-up we have used, in order that the resulting joint
detection probabilities for the initial state (\ref{ME}) do satisfy
all four conditions (\ref{H1})-(\ref{H4}). The basic step towards this
end is to introduce a partially absorbing material into one of the
paths available to either one of the photons (say, photon 2), so that,
actually, {\em only\/} a certain subensemble of the whole ensemble of
emitted photon pairs is analysed for coincidences (see below). This
subensemble will consist of all those pairs of photons 1 and 2 for
which photon 2 is not absorbed and so ends up either in detector $L_2$
or $U_2$ (photon 1 always ends either in $L_1$ or $U_1$). In the
following we shall assume that the phase shifter in beam $B$, besides
imparting a variable phase shift $\phi_2$ to it, multiplies the
amplitude of that beam by $u$ (with $0\leq u \leq 1$). More precisely,
$u$ is the probability amplitude that a photon striking into this
phase shifter passes straight through it, while $v$ is the probability
amplitude that the photon gets absorbed (or, more generally,
scattered), hence $u^2+v^2=1$. Therefore, the state of a photon
propagating along path $B$ will evolve upon interacting with the phase
shifter as
\begin{equation}
\left| B \right\rangle _{2} \rightarrow i\left( ue^{i\phi_2} \left| L_2 \right\rangle + v \left|PS^{\ast} \right\rangle \right) ,
\label{TRANS}
\end{equation}
where the factor $i$ arises from reflection at $M_B$, $\left|
L_2\right\rangle$ denotes the photon transmitted towards detector
$L_2$, and $\left|PS^{\ast} \right\rangle$ denotes an excited state of
the phase shifter due to the absortion of the photon (of course, the
amplitude of absortion $v$ is generally complex, although this is not
relevant for our purpose).

Another necessary modification with respect to the standard
experimental set-up is that the beam splitter $H_i$, $i=1,2$, is no
longer assumed to be \mbox{50-50} (for simplicity, we suppose that it
continues to be symmetric), so that its reflectivity $r_i$ and
transmittivity $t_i$ are any positive real numbers satisfying the
relation $r_i^2 + t_i^2 = 1$. So, any given experimental configuration
for our interferometer set-up is now defined by the values of the
local parameters $\phi_1$, $r_1$ (or $t_1$), $\phi_2$, $r_2$ (or
$t_2$), and $u$. In particular, the four configurations involved in
Hardy's nonlocality argument are determined by the respective
settings: $(\Phi_1,\Phi_2)$, $(\Phi_1,\Phi^{\prime}_2)$,
$(\Phi^{\prime}_1,\Phi_2)$, and $(\Phi^{\prime}_1,\Phi^{\prime}_2)$,
where the arguments $\Phi_1$, $\Phi^{\prime}_1$, $\Phi_2$, and
$\Phi^{\prime}_2$ are a shorthand for the parameters ($\phi_1,
\,r_1$), ($\phi^{\prime}_1, \,r^{\prime}_1$), ($\phi_2, \,r_2, \,u$),
and ($\phi^{\prime}_2, \,r^{\prime}_2, \,u^{\prime}$),
respectively. With all these ingredients, and taking into account the
transformation (\ref{TRANS}), we can calculate again the joint
probabilities in Eq.\ (\ref{P1}) for the initial state
(\ref{ME}). These are given by
\begin{align}
P(L_{1},\Phi_{1};L_{2},\Phi_{2}) &= \tfrac{1}{2} \left[ 
u^{2}r_{1}^{2}t_{2}^{2}+t_{1}^{2}r_{2}^{2}+2ur_{1}r_{2}t_{1}t_{2} 
\cos (\phi_{1}-\phi _{2})\right] , \tag{6a}  \label{PM1}   \\
P(U_{1},\Phi_{1};U_{2},\Phi_{2}) &= \tfrac{1}{2} \left[
u^{2}t_{1}^{2}r_{2}^{2}+r_{1}^{2}t_{2}^{2}+2ur_{1}r_{2}t_{1}t_{2} 
\cos (\phi_{1}-\phi _{2})\right] .  \tag{6b}  \label{PM2}
\end{align}
Of course, in the case in which $r_i=t_i=1/{\sqrt 2}$, and $u=1$, we retrieve the probabilities in Eq.\ (\ref{P1}). Hardy's nonlocality conditions are now defined by
\begin{align}
P(L_{1},\Phi _{1};L_{2},\Phi _{2}) & = 0\, , \tag{7a} \label{HM1}  \\
P(U_{1},\Phi _{1};U_{2},\Phi _{2}^{\prime }) & = 0\, , \tag{7b}  \label{HM2}  \\
P(U_{1},\Phi _{1}^{\prime };U_{2},\Phi _{2}) & = 0\, , \tag{7c}  \label{HM3}  \\
P(U_{1},\Phi _{1}^{\prime };U_{2},\Phi _{2}^{\prime }) & > 0\, . \tag{7d}  \label{HM4}
\end{align}
So, in view of Eqs.\ (\ref{PM1}) and (\ref{PM2}), we must have the following relations in order for the conditions (\ref{HM1}), (\ref{HM2}), and (\ref{HM3}), respectively, to be satisfied by the state (\ref{ME})
\begin{align}
\cos (\phi _{1}-\phi _{2}) &= - \frac{%
u^{2}r_{1}^{2}t_{2}^{2}+t_{1}^{2}r_{2}^{2}}{2ur_{1}r_{2}t_{1}t_{2}},\tag{8a} \label{C1}  \\
\cos (\phi _{1}-\phi _{2}^{\prime }) &= - \frac{(u^{\prime})
^{2}t_{1}^{2}(r_{2}^{\prime })^{2}+r_{1}^{2}(t_{2}^{\prime })^{2}}{%
2u^{\prime }r_{1}r_{2}^{\prime }t_{1}t_{2}^{\prime }},  \tag{8b} \label{C2} \\
\cos (\phi _{1}^{\prime }-\phi _{2}) &= - \frac{u^{2}(t_{1}^{\prime
})^{2}r_{2}^{2}+(r_{1}^{\prime })^{2}t_{2}^{2}}{2ur_{1}^{\prime
}r_{2}t_{1}^{\prime }t_{2}}. \tag{8c}  \label{C3}
\setcounter{equation}{8}
\end{align}

It is straightforward to see that the right-hand side for each of the
Eqs.\ (\ref{C1})-(\ref{C3}) has the value $-1$ as its upper
bound. Therefore, the only way to satisfy the conditions
(\ref{HM1})-(\ref{HM3}) is to choose the parameters in the right-hand
sides of Eqs.\ (\ref{C1})-(\ref{C3}) such that each of these sides
equals $-1$. The necessary and sufficient condition in order for the
right-hand side of Eqs.\ (\ref{C1}), (\ref{C2}), and (\ref{C3}) to be
equal to $-1$ is, respectively,
\begin{align}
ur_1 t_2 &= t_1 r_2 \, , \tag{9a} \label{CO1}  \\
u^{\prime}t_1 r^{\prime}_2 &= r_1 t^{\prime}_2 \, , \tag{9b} \label{CO2}   \\
ut^{\prime}_1 r_2 &= r^{\prime}_1 t_2 \, . \tag{9c} \label{CO3}
\setcounter{equation}{9}
\end{align}
The set of equations (\ref{CO1})-(\ref{CO3}) admits an infinite number
of solutions. So, we shall assume that the parameters $r_i$, $t_i$,
$r^{\prime}_i$, $t^{\prime}_i$, $u$, and $u^{\prime}$ have been chosen
such that they satisfy that set of equations. An immediate consequence
of Eqs.\ (\ref{CO1})-(\ref{CO3}) is
\begin{equation}
u^2 u^{\prime} t^{\prime}_1 r^{\prime}_2 = r^{\prime}_1 t^{\prime}_2 \, . \label{CO4}
\end{equation}
On the other hand, since $\cos (\phi _{1}-\phi _{2})=-1$, $\cos (\phi
_{1}-\phi _{2}^{\prime })=-1$, and \linebreak $\cos (\phi_{1}^{\prime
}-\phi _{2})=-1$ (see Eqs.\ (\ref{C1})-(\ref{C3}) and
(\ref{CO1})-(\ref{CO3})), we have necessarily $\cos (\phi_{1}^{\prime
}-\phi_{2}^{\prime})=-1$. Thus, taking into account this latter
equality, and the relation (\ref{CO4}), we arrive at the following
expression for the probability in Eq.\ (\ref{HM4})
\begin{equation}
P(U_{1},\Phi _{1}^{\prime };U_{2},\Phi _{2}^{\prime }) = \frac{1}{2} 
\left[ u^{\prime} t_{1}^{\prime} r_{2}^{\prime} (1-u^2 ) \right]^2.  \label{P}
\end{equation}
The value of the probability (\ref{P}) is a direct measure of the
degree of nonlocality inherent in the Hardy equations
(\ref{HM1})-(\ref{HM4}). It is obvious that, for given
$t_{1}^{\prime}$, $r_{2}^{\prime}$, and $u$, the maximum value of
(\ref{P}) is attained for the choice $u^{\prime}=1$. So, unless
otherwise stated, we shall throughout fix the parameter $u^{\prime}$
to be unity.

A few remarks concerning the probability (\ref{P}) should be added
here. In the first place, we can see that $P(U_{1},\Phi_{1}^{\prime
};U_{2},\Phi_{2}^{\prime })$ vanishes whenever $u^2 =1$.\footnote{%
Analogously, the probability (27) in Ref.\ \cite{Wu et al.} vanishes
whenever the parameter $\beta$ is unity (see also Ref.\
\cite{Cereceda97a}).} This fact conforms to the analysis made in the
preceding section according to which no Hardy-type contradiction for
the maximally entangled state is possible if each particle is
subjected to a dichotomic choice. Therefore, in order for the state
(\ref{ME}) to satisfy all four conditions (\ref{HM1})-(\ref{HM4}),
there must be a {\em nonzero\/} probability that photon 2 in beam $B$
gets absorbed inside the phase shifter when the experimental set-up is
arranged to be in the configuration $(\Phi_1,\Phi_2)$ or
$(\Phi^{\prime}_1,\Phi_2)$. Accordingly, for these configurations,
only a statistical fraction $u^2$ (with $u<1$) of the ensemble of
emitted photons following path $B$ will reach detectors $L_2$ or
$U_2$. On the other hand, it should be noticed that the probability
function (\ref{P}) (with $u^{\prime}=1$) involves only two independent
variables as, from Eq.\ (\ref{CO4}), the parameter $u$ is constrained
to obey the relation
\begin{equation}
u^2 = \frac{r^{\prime}_1 t^{\prime}_2}{t^{\prime}_1 r^{\prime}_2}
= \frac{\sqrt{ [1-(t^{\prime}_1)^{2}] [1-(r^{\prime}_2)^{2}]}}
{t^{\prime}_1 r^{\prime}_2}.   \label{u}
\end{equation}
Therefore, by inserting the expression (\ref{u}) into Eq.\ (\ref{P}),
we make the probability $P(U_{1},\Phi_{1}^{\prime
};U_{2},\Phi_{2}^{\prime })$ to depend on the parameters
$t^{\prime}_1$ and $r^{\prime}_2$. It is important to realize,
however, that, although $t^{\prime}_1$ and $r^{\prime}_2$ are two
independent variables with range of variation $0 \leq
t^{\prime}_1,r^{\prime}_2 \leq 1$, {\it not} all the combinations of
values $(t^{\prime}_1, r^{\prime}_2)$ are to be permitted, if we want
the squared parameter $u^2$ in Eq.\ (\ref{u}) to represent a physical
probability. Specifically, the allowed values of $t^{\prime}_1$ and
$r^{\prime}_2$ are those for which the quotient in Eq.\ (\ref{u}) is
less than (or equal to) unity. One can easily verify that the values
of $t^{\prime}_1$ and $r^{\prime}_2$ for which $u^2 \leq 1$ are those
satisfying
\begin{equation}
( t^{\prime}_1 )^2 + ( r^{\prime}_2 )^2 \geq 1 ,
\label{GEQ}
\end{equation}
with $u^2 =1$ for $(t^{\prime}_1 )^2 + (r^{\prime}_2 )^2 =1$, and $u^2
< 1$ for $(t^{\prime}_1 )^2 + (r^{\prime}_2 )^2 > 1$. Note that the
pair of values $( t^{\prime}_1 = 1/{\sqrt 2}, r^{\prime}_2 = 1/{\sqrt
2})$ fulfills the equality in Eq.\ (\ref{GEQ}), and then, for such
values, $P(U_{1},\Phi_{1}^{\prime };U_{2},\Phi_{2}^{\prime
})=0$. Hence, in order for all the equations (\ref{HM1})-(\ref{HM4})
to be satisfied for the initial state (\ref{ME}), either beam splitter
$H^{\prime}_1$ or $H^{\prime}_2$ should not be 50-50. It will further
be noted that the remaining parameters $r_1$ and $r_2$ (recall that
$t_i =\sqrt{1- r_i^2}$) turn out to be determined by the values of
$t^{\prime}_1$ and $r^{\prime}_2$. Indeed, by Eq.\ (\ref{CO2}) (with
$u^{\prime} =1$) and Eq.\ (\ref{CO3}), we have, respectively, $r_1
/t_1 = r^{\prime}_2 /t^{\prime}_2$ and $t_2 /r_2 = u t^{\prime}_1
/r^{\prime}_1$. Regarding the phases $\phi_1$, $\phi^{\prime}_1$,
$\phi_2$, and $\phi^{\prime}_2$, we can choose one of them in an
unrestricted way, while the three other are forced to accommodate. So,
for instance, suppose $\phi^{\prime}_2$ is set to $\phi_0$, with
$\phi_0$ taking on any {\it arbitrary} value. Then, as $\phi_1 -
\phi_2 = n_1\pi$, $\phi_1 - \phi_0 = n_2\pi$, and $\phi^{\prime}_1 -
\phi_2 = n_3\pi$ (with $n_1$, $n_2ç$, and $n_3$ being odd integers),
the remaining phases $\phi_1$, $\phi_2$, and $\phi^{\prime}_1$, are
fixed to be $n_2 \pi+\phi_0$, $(n_2 - n_1)\pi +\phi_0$, and $(n_3 +
n_2 - n_1)\pi +\phi_0$, respectively. On the other hand, for the
particular case in which $t^{\prime}_1 = r^{\prime}_2 \equiv q$,
$1/{\sqrt 2} \leq q \leq 1$, Eq.\ (\ref{P}) (with $u^{\prime}=1$)
reduces to
\begin{equation}
P(U_{1},\Phi _{1}^{\prime };U_{2},\Phi _{2}^{\prime }) = \frac{1}{2}
\left( 2q^2 - 1 \right) ^2.  \label{PP}
\end{equation}
It is worth noting, incidentally, that, whenever $t^{\prime}_1 =
r^{\prime}_2 \equiv q$, the beam splitter $H_2$ turns out to be 50-50,
irrespective of the value taken by $q$. Indeed, since $u^2 =
\frac{1}{q^2} -1$ for this case, we find that the expression
$(r_2/t_2)^2 = u^{-2}(\frac{1}{q^2} -1)$ is, identically,
unity. Eq. (\ref{PP}) has a maximum value of $\frac{1}{2}$ when
$q=1$. Clearly, this extremum corresponds to the maximum absolute
value attainable by the probability (\ref{P}), which is achieved for
$u^{\prime}=1$, $u=0$, and $t^{\prime}_1 = r^{\prime}_2 =1$.

Suppose now that the two-photon interferometer is operating in the
initial configuration $(\Phi_1,\Phi_2)$. Then we can use predictions
(\ref{HM1})-(\ref{HM4}) to obtain a probabilistic contradiction with
the assumption of local realism, provided we restrict the analysis to
the subensemble of joint detection events for which photon 1 ends
either in $L_1$ or $U_1$ {\em and\/} photon 2 ends either in $L_2$ or
$U_2$. So, from Eq.\ (\ref{HM1}), we can quickly deduce that, (i)
either one or both of photons 1 and 2 in a pair pertaining to this
selected subensemble must have reached the corresponding $U$-detector,
that is, we must have either $U_1 =1$ or $U_2 =1$ (or else $U_1 = U_2
=1$), where the notation $U_i =1$ is used to indicate that a count is
registered by detector $U_i$. On the other hand, from Eq.\
(\ref{HM2}), and applying the notion of local realism, we can infer
that, (ii) if $U_1 =1$, then we would have obtained the null result
$U_2 =0$ if the experimental configuration had been set to
$(\Phi_1,\Phi^{\prime}_2)$, instead of $(\Phi_1,\Phi_2)$, where the
notation $U_i =0$ is meant to signify the very absence of a count
triggering detector $U_i$. Similarly, from Eq.\ (\ref{HM3}), and
according to local realism, we can conclude that, (iii) if $U_2 =1$,
then we would have obtained $U_1 =0$, had the configuration been set
to $(\Phi^{\prime}_1,\Phi_2)$. Therefore, from results (i)-(iii), and
applying once more the assumption of local realism, it follows that,
if the interferometer set-up had been arranged to operate in the
configuration $(\Phi^{\prime}_1, \Phi^{\prime}_2)$, then at least one
of the two photons 1 or 2 in a pair pertaining to the above-defined
subensemble would {\em not\/} have impinged on the corresponding
$U$-detector (that is, $U_1 U_2 =0$), in contradiction with the
quantum prediction in Eq.\ (\ref{HM4}) which allows the possibility
that {\em both\/} photons of any emitted pair end in the corresponding
$U$-detector for the configuration $(\Phi^{\prime}_1,
\Phi^{\prime}_2)$ (that is, $U_1 U_2 =1$). The statistical fraction of
emitted photon pairs for which photon 2 reaches either $L_2$ or $U_2$
in the configuration $(\Phi_1,\Phi_2)$ is $\frac{1}{2} (1+u^2)$. This
fraction can be made arbitrarily close to unity by letting $u$ tend to
$1$. It is to be noted, however, that the probability (\ref{HM4})
tends to zero as the parameter $u$ approaches unity. Thus we deduce
that, in order to get a finite probability of contradiction with the
assumption of local realism, one must necessarily consider a {\em
proper\/} subensemble of the total ensemble of emitted photon
pairs. In addition to this, it should be noticed that the above
argument for nonlocality does not hinge on the choice $u^{\prime} =1$
but works for any value of $u^{\prime}$ different from zero (the value
$u^{\prime} = 0$ is excluded since, for this value, the probability
(\ref{HM4}) is zero, as can be seen from Eq.\ (\ref{P})). This follows
from the fact that the local realistic prediction (ii) above is a {\em
negative\/} statement in the sense that it says nothing about which
detector photon 2 will actually hit. Instead, it tells us where photon
2 will {\em not\/} be found. Of course, if $U_2 =0$ and $u^{\prime}
<1$, photon 2 can still end either in detector $L_2$ or inside the
absorber, but the specific location of photon 2 is not relevant to the
argument. The important thing to be learned from prediction (ii) is
that photon 2 will not be detected by $U_2$. Applying this kind of
prediction (specifically, predictions (ii) and (iii) above) to a
certain subensemble of postselected photon pairs leads us ultimately
to a contradiction with the quantum prediction in Eq.\ (\ref{HM4}).

As is known \cite{Mermin,Hardy94}, Hardy's nonlocality argument can equally be cast in the form of a simple inequality involving the four probabilities in Eqs.\ (\ref{HM1})-(\ref{HM4}):
\begin{equation}
\negthickspace  \negthickspace
\negthickspace  \negthickspace
\negthickspace  \negthickspace
\negthickspace  
P(U_{1},\Phi _{1}^{\prime };U_{2},\Phi _{2}^{\prime }) \leq
P(L_{1},\Phi _{1};L_{2},\Phi _{2}) + P(U_{1},\Phi _{1};U_{2},\Phi _{2}^{\prime })
+ P(U_{1},\Phi _{1}^{\prime };U_{2},\Phi _{2}) ,  \label{CH}
\end{equation}
which is a particular case of the CH inequality \cite{C-H,C-S}. For
the interferometric set-up devised in this section, the inequality
(\ref{CH}) is maximally violated for the values $P(L_{1},\Phi
_{1};L_{2},\Phi _{2}) = P(U_{1},\Phi _{1};U_{2},\Phi _{2}^{\prime }) =
P(U_{1},\Phi _{1}^{\prime };U_{2},\Phi _{2}) =0$, and $P(U_{1},\Phi
_{1}^{\prime };U_{2},\Phi _{2}^{\prime }) = 1/2$. So our proof gives a
probability of contradiction with local realism of up to $50\%$, which
substantially improves on the maximum probability of contradiction
($\approx 9\%$) obtained in the standard Hardy's proof for
less-than-maximally entangled states \cite{Hardy93}. It should be
emphasized, however, that, as far as the above nonlocality proof is
concerned, only a restricted ensemble of joint detection events has
been used to get the contradiction. In the next section we shall
derive a more general Bell inequality which refers to the total set of
localization correlations, without any selection of any
subsensemble. As we shall see, this latter inequality is not violated
in any case by the quantum-mechanical predictions.

\section{Bell-CH inequality for the entire pattern of correlations}

We now deduce the Bell inequality that obtains when all the
coincidence detection events are taken into account in the statistical
analysis, including those events for which photon 2 gets absorbed
inside the phase shifter. In this section we relax the condition
$u^{\prime}=1$, so that, as was already implicitly assumed at the end
of the preceding section when we developed the argument for
nonlocality, it is supposed that $u^{\prime}$ may take on any value
different from zero. Let us consider the Clauser-Horne formulation
\cite{C-H,C-S} of the probabilistic approach to local realism. The
notion of realism is introduced by assuming that some hidden variables
$\lambda$ exist that represent the complete physical state of each
individual pair of correlated photons emanating from the
source. Within this probabilistic approach, the hidden variable
description does not uniquely determine the outcome of any given
measurement but only determines the respective probabilities for the
various possible outcomes that may occur in a given measurement. So,
for example, $P_{\lambda}(U_1,\Phi_1^{\prime})$ is the probability of
detecting photon 1 in $U_1$, given the state $\lambda$ of the
individual pair of photons and the parameters ($\phi^{\prime}_1,
\,r^{\prime}_1$) of the `outer' interferometer with arms $A$ and $D$
(see Fig.~1); $P_{\lambda}(U_2,\Phi_2^{\prime})$ is the probability
for photon 2 to end in $U_2$, given $\lambda$ and the parameters
($\phi^{\prime}_2, \,r^{\prime}_2, \,u^{\prime}$) of the `inner'
interferometer with arms $B$ and $C$; and $P_{\lambda}(U_{1},\Phi
_{1}^{\prime };U_{2},\Phi _{2}^{\prime })$ is the joint probability
that photons 1 and 2 in the state $\lambda$ are detected by $U_1$ and
$U_2$, respectively, when the experimental configuration is set to
$(\Phi^{\prime}_1,\Phi^{\prime}_2)$. The assumption of locality, on
the other hand, is expressed by the following factorizability
condition\footnote{The locality condition in Eq.\ (\ref{factor}) can
be viewed as the logical conjunction of two assumptions sometimes
referred to as `parameter independence' and `outcome
independence'. For a detailed account of this subject see Refs.\
\cite{Jarrett,Shimony} and Appendix B of Ref.\ \cite{Mermin}.}
\begin{equation}
P_{\lambda}(U_{1},\Phi _{1}^{\prime };U_{2},\Phi _{2}^{\prime }) =
P_{\lambda}(U_1,\Phi_1^{\prime}) P_{\lambda}(U_2,\Phi_2^{\prime}).  \label{factor}
\end{equation}
The ensemble (observable) probability $P(U_{1},\Phi _{1}^{\prime
};U_{2},\Phi _{2}^{\prime })$ of jointly obtaining $U_1 =1$ and $U_2
=1$ for the configuration $(\Phi^{\prime}_1,\Phi^{\prime}_2)$ is
expressible as a weighted average of the individual probabilities
\begin{equation}
P(U_{1},\Phi _{1}^{\prime };U_{2},\Phi _{2}^{\prime }) = \langle
P_{\lambda}(U_{1},\Phi _{1}^{\prime };U_{2},\Phi _{2}^{\prime })
\rangle = \langle P_{\lambda}(U_1,\Phi_1^{\prime})
P_{\lambda}(U_2,\Phi_2^{\prime}) \rangle .  \label{enspro}
\end{equation}
It is further assumed that the underlying normalised probability
distribution $\rho (\lambda)$ corresponding to the initial state of
the emitted photon pairs as well as the set $\Lambda$ of values of
$\lambda$ are independent of the actual configuration of the
experimental set-up.

When the configuration is $(\Phi_1,\Phi_2)$, photon 1 ends in either
$L_1$ or $U_1$, whereas photon 2 ends in either $L_2$, $U_2$, or $A_2$
(where $A_2$ denotes the absorbing phase shifter acting on beam
$B$). Therefore, the individual probabilities should satisfy the
following relations
\begin{align}
P_{\lambda}(L_1,\Phi_1)& + P_{\lambda}(U_1,\Phi_1) = 1 , \tag{18a}
\label{rel1} \\ P_{\lambda}(L_2,\Phi_2)& + P_{\lambda}(U_2,\Phi_2) +
P_{\lambda}(A_2,\Phi_2) = 1 .  \tag{18b} \label{rel2}
\setcounter{equation}{18}
\end{align}
Thus the joint probability (\ref{enspro}) can equivalently be written as
\begin{eqnarray}
P(U_{1},\Phi _{1}^{\prime }; U_{2},\Phi _{2}^{\prime }) & & \nonumber
\\ = \langle P_{\lambda} (U_1, \Phi_1^{\prime}) & \!\!\!\!\!\!\!\!\!\!
P_{\lambda}(L_1,\Phi_1) \, P_{\lambda}(L_2,\Phi_2) \,
P_{\lambda}(U_2,\Phi_2^{\prime}) \rangle & \nonumber \\ + \; \langle
P_{\lambda}(U_1, & \Phi_1^{\prime})\, P_{\lambda}(U_1,\Phi_1) \,
P_{\lambda}(L_2,\Phi_2) \, P_{\lambda}(U_2,\Phi_2^{\prime}) \rangle &
\nonumber \\ + \; \langle P_{\lambda}(U_1, & \Phi_1^{\prime})\,
P_{\lambda}(L_1,\Phi_1)\, P_{\lambda}(U_2,\Phi_2)
\,P_{\lambda}(U_2,\Phi_2^{\prime}) \rangle & \nonumber \\ + \; \langle
P_{\lambda}(U_1, & \Phi_1^{\prime})\, P_{\lambda}(U_1,\Phi_1)\,
P_{\lambda}(U_2,\Phi_2)\, P_{\lambda}(U_2,\Phi_2^{\prime}) \rangle &
\nonumber \\ + \; \langle P_{\lambda}(U_1, & \Phi_1^{\prime})\,
P_{\lambda}(L_1,\Phi_1)\, P_{\lambda}(A_2,\Phi_2)\,
P_{\lambda}(U_2,\Phi_2^{\prime}) \rangle & \nonumber \\ + \; \langle
P_{\lambda}(U_1, & \Phi_1^{\prime})\, P_{\lambda}(U_1,\Phi_1)\,
P_{\lambda}(A_2,\Phi_2)\, P_{\lambda}(U_2,\Phi_2^{\prime}) \rangle &
. \label{six}
\end{eqnarray}
Each of the factors $P_{\lambda}(\cdots)$ inside the angular brackets
$\langle \; \rangle$ appearing in Eq.\ (\ref{six}) is nonnegative and
fulfills the relation $P_{\lambda}(\cdots) \leq 1$. So it is evident
that the average values on the right of Eq.\ (\ref{six}) can be
bounded as follows \cite{Mermin}
\begin{align}
\langle P_{\lambda} (U_1, \Phi_1^{\prime})\, & P_{\lambda}(L_1,\Phi_1)
\, P_{\lambda}(L_2,\Phi_2) \, P_{\lambda}(U_2,\Phi_2^{\prime}) \rangle
\nonumber \\ & \leq \langle P_{\lambda}(L_1,\Phi_1) \,
P_{\lambda}(L_2,\Phi_2) \rangle = P(L_{1},\Phi _{1};L_{2},\Phi _{2})
\, , \tag{20a} \label{des1}
\end{align}
\begin{align}
\langle P_{\lambda} (U_1, \Phi_1^{\prime})\, & P_{\lambda}(U_1,\Phi_1)
\, P_{\lambda}(L_2,\Phi_2) \, P_{\lambda}(U_2,\Phi_2^{\prime}) \rangle
\nonumber \\ & \leq \langle P_{\lambda}(U_1,\Phi_1) \,
P_{\lambda}(U_2,\Phi_2^{\prime}) \rangle = P(U_{1},\Phi
_{1};U_{2},\Phi _{2}^{\prime }) \, \tag{20b} ,
\end{align}
\begin{align}
\langle P_{\lambda} & (U_1, \Phi_1^{\prime}) \, P_{\lambda}
(L_1,\Phi_1) \, P_{\lambda}(U_2,\Phi_2) \,
P_{\lambda}(U_2,\Phi_2^{\prime}) \rangle \nonumber \\ & + \langle
P_{\lambda} (U_1, \Phi_1^{\prime}) \, P_{\lambda}(U_1,\Phi_1) \,
P_{\lambda} (U_2,\Phi_2) \, P_{\lambda}(U_2,\Phi_2^{\prime}) \rangle
\nonumber \\ & \qquad = \, \langle P_{\lambda} (U_1, \Phi_1^{\prime})
\, P_{\lambda}(U_2,\Phi_2) \, P_{\lambda}(U_2,\Phi_2^{\prime}) \rangle
\nonumber \\ & \qquad \qquad \leq \langle
P_{\lambda}(U_1,\Phi_1^{\prime}) \, P_{\lambda}(U_2,\Phi_2) \rangle =
P(U_{1},\Phi_{1}^{\prime};U_{2},\Phi _{2}) \, , \tag{20c}
\end{align}
and
\begin{align}
\langle P_{\lambda} & (U_1, \Phi_1^{\prime}) \, P_{\lambda}
(L_1,\Phi_1) \, P_{\lambda}(A_2,\Phi_2) \,
P_{\lambda}(U_2,\Phi_2^{\prime}) \rangle \nonumber \\ & + \langle
P_{\lambda} (U_1, \Phi_1^{\prime}) \, P_{\lambda}(U_1,\Phi_1) \,
P_{\lambda} (A_2,\Phi_2) \, P_{\lambda}(U_2,\Phi_2^{\prime}) \rangle
\nonumber \\ & \qquad \leq \langle P_{\lambda} (L_1,\Phi_1) \,
P_{\lambda}(A_2,\Phi_2) \rangle + \langle P_{\lambda}(U_1,\Phi_1) \,
P_{\lambda} (A_2,\Phi_2) \rangle \nonumber \\ & \qquad \qquad =
\langle P_{\lambda}(A_2,\Phi_2) \rangle = P(A_{2}, \Phi _{2}) \, ,
\tag{20d} \label{des4} \setcounter{equation}{20}
\end{align}
where $P(A_{2},\Phi _{2})= P(L_1,\Phi_{1};A_2,\Phi_2)
+P(U_1,\Phi_{1};A_2,\Phi_2)$ corresponds to the probability that
photon 2 is absorbed by the phase shifter, given the configuration
$\Phi_2$ for the inner interferometer. From Eqs.\ (\ref{six}) and
(\ref{des1})-(\ref{des4}), we finally obtain
\begin{eqnarray}
P(U_{1},\Phi _{1}^{\prime };U_{2},\Phi _{2}^{\prime }) \leq & \,
P & (L_{1},\Phi _{1};L_{2},\Phi _{2}) + P(U_{1},\Phi _{1};U_{2},\Phi _{2}^{\prime })
\nonumber  \\
& \, + &  P(U_{1},\Phi _{1}^{\prime };U_{2},\Phi _{2}) + P(A_{2},\Phi _{2}) \, .  \label{CH-total}
\end{eqnarray}
Inequality (\ref{CH-total}) should be satisfied by any local realistic
theory, and it is applicable in the case that the entire pattern of
localization correlations is analysed. In the case where {\em only\/}
those events for which photon 2 reaches either $L_2$ or $U_2$ are
considered (with the actual configuration of the inner interferometer
being $\Phi_2$), we may dispense with the last term on the right of
Eq.\ (\ref{CH-total}), and then the above inequality (\ref{CH-total})
reduces to the inequality (\ref{CH}). On the other hand, when the
Hardy nonlocality conditions in Eqs.\ (\ref{HM1})-(\ref{HM4}) are
fulfilled, the inequality (\ref{CH-total}) simplifies to
\begin{equation}
P(U_{1},\Phi _{1}^{\prime };U_{2},\Phi _{2}^{\prime }) \, \leq \,
P(A_{2},\Phi _{2}) .
\label{CH-sim}
\end{equation}
It is straightforward to see that the quantum-mechanical predictions
always satisfy the inequality (\ref{CH-sim}). Indeed, recalling the
quantum prediction (\ref{P}), and taking into account that
$P(A_{2},\Phi _{2}) = \frac{1}{2} (1-u^2)$, inequality (\ref{CH-sim})
reads as (provided $u^2 \neq 1$)
\begin{equation}
\left( u^{\prime} t_{1}^{\prime} r_{2}^{\prime} \right)^2 (1-u^2 )
\, \leq \, 1.    \label{CH-final}
\end{equation}
Obviously, inequality (\ref{CH-final}) is satisfied for any values of
$u^{\prime}$, $t_{1}^{\prime}$, $r_{2}^{\prime}$, and $u$. (Of course,
as we know, the variables $u^{\prime}$, $t_{1}^{\prime}$,
$r_{2}^{\prime}$, and $u$ are not all independent. They must fulfill
the relation (\ref{CO4}) if the conditions (\ref{HM1})-(\ref{HM3}) are
to be satisfied.)

In view of the fulfillment of the relevant inequality (\ref{CH-sim})
by the quantum-mechanical predictions, one might wonder whether some
other Bell-type inequality might be violated for the same
situation. To answer this question we note that the inequality
(\ref{CH-total}) is completely general in the sense that no assumption
other than local realism has been used in its derivation. Therefore,
for the situation in which the Hardy conditions
(\ref{HM1})-(\ref{HM4}) hold for the maximally entangled state, we
argue that no Bell-type inequality exists that is violated by the
predictions of quantum mechanics, provided the total set of
localization correlations is considered. Moreover, the inequality in
Eq.\ (\ref{CH-total}) appears to be the most obvious one for
comparison with earlier work on the subject.

\section{Interaction-free measurement in the modified interferometer set-up}

As already emphasized, the parameter $u$ must be strictly less than
unity if we want the initial state (\ref{ME}) to satisfy all the Hardy
conditions (\ref{HM1})-(\ref{HM4}). So there must be a nonvanishing
probability for a photon traveling path $B$ to be absorbed by the
phase shifter (which, in what follows, will be referred to as the
`object') in either configuration $(\Phi_1,\Phi_2)$ or
$(\Phi^{\prime}_1,\Phi_2)$. We now show how our interferometer set-up
can be used to detect the presence of the (in general partially)
absorbing object in path $B$, without any photon being scattered from
it. Actually, our method generalises the original proposal of Elitzur
and Vaidman (EV) who demonstrate the principle of an interaction-free
measurement (IFM) by placing a perfect absorber in one arm of a
standard Mach-Zehnder one-particle interferometer \cite{EV,Vaidman}. A
previous extension of EV's original scheme from single-particle to
two-particle case has been recently carried out by Noh and Hong
\cite{Noh-Hong} who developed a new scheme of IFM that is based on
nonclassical fourth-order (i.e. two-photon) interference
effect. Although the IFM scheme presented here is based on this same
effect, it differs from that of Ref.\ \cite{Noh-Hong} in that the
latter utilizes a {\em single\/} Mach-Zehnder type two-photon
interferometer, whereas the former utilizes {\em two\/} Mach-Zehnder
type one-photon interferometers (see Fig.\ 1). As we shall see, our
IFM scheme gives a maximum fraction of successful (interaction-free)
measurements greater than that obtained in Ref.\ \cite{Noh-Hong}.

In the original EV scheme the interferometer is arranged so that, due
to destructive interference, no photon is detected at one of the two
output ports (the `dark' port). Blocking one of the two arms of the
interferometer destroys the interference and then some of the photons
will reach the dark output port, thus indicating that something stands
in one of the two possible paths inside the interferometer
\cite{EV}. Likewise, as a matter of fact, the insertion of a partially
absorbing object in one of the paths (say path $B$) of our
interferometer set-up, modifies the {\em two-photon\/} interference
pattern obtained when no absorbing material is present. Therefore, the
observation of a previously forbidden two-photon coincidence count
would entail an interaction-free measurement of the presence of the
object. To see this, let us consider, for example, the configuration
$(\Phi_1^{\prime},\Phi_2)$ with the phases $\phi_1^{\prime}$ and
$\phi_2$ fulfilling $\phi_1^{\prime} - \phi_2 = n_3 \pi$ ($n_3 =
\pm1,\pm3,\ldots$\,). Then, from Eq.\ (\ref{PM1}), we have
\begin{equation}
P(L_{1},\Phi_{1}^{\prime};L_{2},\Phi_{2}) = \frac{1}{2} \left( 
u r_{1}^{\prime} t_{2} - t_{1}^{\prime} r_{2}  \right) ^2 . \label{IF1}  
\end{equation}
Now, if the parameters $u$, $r_1^{\prime}$, $t_2$, $t_1^{\prime}$, and
$r_2$ are constrained to obey the relation (\ref{CO3}), we can readily
express the joint detection probability (\ref{IF1}) as a function of
the two parameters $u$ and $r_2$ as follows
\begin{equation}
P(L_{1},\Phi_{1}^{\prime};L_{2},\Phi_{2}) = \frac{1}{2} \left[ 
\frac{ (1- r_2^2) r_2^2 (1-u^2) ^2} {1 - r_2^2 (1- u^2)}
\right] .  \label{IF2}
\end{equation}
From Eq.\ (\ref{IF2}) we see that
$P(L_{1},\Phi_{1}^{\prime};L_{2},\Phi_{2}) = 0$ whenever $u^2 =1$
(that is, for a perfectly transmitting object), whereas
$P(L_{1},\Phi_{1}^{\prime};L_{2},\Phi_{2}) > 0$ whenever $u^2 <1$ and
$r_2 \neq 0, \,1$. So, when a partially absorbing object is introduced
in path $B$, there will be a chance for photon 1 to be registered by
detector $L_1$ and, at the same time, for photon 2 to be registered by
$L_2$. Thus, whenever a coincidence registration in detectors $L_1$
and $L_2$ is observed, one can deduce that a partially absorbing
object is certainly present in one of routes available to the photons
inside the interferometer, without actually any photon having been
scattered by the object. The maximum probability for this coincidence
detection event to occur is 1/2. The probability (\ref{IF2}) tends to
the upper limit of 1/2 when $u=0$ and $r_2 \rightarrow 1$, that is,
for a {\em perfectly\/} absorbing object and for an almost perfectly
reflecting beam splitter $H_2$. Under these conditions (namely, $u=0$
and $r_2 \rightarrow 1$) the reflectivity of beam splitter
$H_1^{\prime}$ turns out to be equal to zero (as $r_1^{\prime}
/t_1^{\prime} = u r_2 /t_2$). On the other hand, when $u=0$, the
probability for photon 2 to be absorbed by the object is 1/2 since,
for the state (\ref{ME}), photon 2 enters beam $B$ with probability
1/2. So, the maximum fraction of measurements that can be
interaction-free in the present IFM procedure is
\begin{equation}
\eta_{\mathrm{max}} = \frac{P_{\mathrm{max}}(L_{1},\Phi_{1}^{\prime};L_{2},\Phi_{2})}
{P_{\mathrm{max}}(L_{1},\Phi_{1}^{\prime};L_{2},\Phi_{2})+ P_{\mathrm{max}}(\mathrm{abs})}
=\frac{1}{2} .
\end{equation}
Thus, for $u=0$ and $r_2 \rightarrow 1$, about half of the
measurements yields conclusive information about the presence of the
object, apparently without interacting with it. This 50\%-efficiency
of the IFM scheme just described is greater than the nominal value
$\eta =\frac{1}{3}$ obtained for the IFM scheme of Ref.\
\cite{Noh-Hong},\footnote{We believe, incidentally, that, actually,
the correct value for the efficiency corresponding to this latter IFM
scheme is $\eta = \frac{1}{4}$. This lowering stems from the fact
that, when evaluating the parameter $\eta$, the authors do not take
into account the unfavourable cases in which {\em both\/} photons of a
pair are detected in either D1 or D2 (see Ref.\ \cite{Noh-Hong} for
the details of the involved set-up).}  although both of them are based
on the same principle, namely nonclassical fourth-order interference.

Furthermore, it is to be mentioned that the insertion of a partially
absorbing object in path $B$ does not change the probabilities of
single detections $P(L_1)$ and $P(U_1)$ by detectors $L_1$ and $U_1$,
respectively. Indeed, it is easily shown that $P(L_1)=P(U_1)=1/2$,
irrespective of the value of $u$. On the other hand, the single
detection probabilities $P(L_2)$ and $P(U_2)$ are found to be
$P(L_2)=\frac{1}{2}(u^2 t_2^2+r_2^2)$ and $P(U_2)=\frac{1}{2}(u^2
r_2^2+t_2^2)$. Obviously, the quantity $1-(P(L_2)+P(U_2))=\frac{1}{2}
(1-u^2)$ corresponds to the probability of photon 2 being absorbed by
the object. For a perfectly absorbing object and for an almost
perfectly reflecting beam splitter $H_2$ we can make $P(L_2)
\rightarrow 1/2$ and $P(U_2) \rightarrow \nolinebreak 0$. However,
even in the extreme case in which the count rate of detector $U_2$
practically vanishes, we cannot conclude at all from the sole
observation of a count at either detector $L_2$ or $U_2$ that an
absorbing object is in place, since the photon could have reached
detector $L_2$ or $U_2$ in both cases: when the object is, or when the
object is not, inserted in path $B$ of the interferometer. This is in
sharp contrast with the situation described in the preceding paragraph
where the observation of a {\em single\/} coincidence count at
detectors $L_1$ and $L_2$ enables one to conclusively determine the
presence of the object without scattering a single photon.

\section{Conclusions and final remarks}

Hardy's original proof of nonlocality \cite{Hardy93,Goldstein94} does
not work for the maximally entangled state. Nevertheless, as we have
shown by using a modified two-particle interferometer set-up, a formal
nonlocality contradiction of the Hardy type can be established for the
maximally entangled state if not all the coincidence detections are
taken into account in the statistical analysis. Within the selected
subensemble, a Bell inequality may be violated. A proof of nonlocality
of this kind was already derived by Wu et al.\ in Ref.\ \cite{Wu et
al.} where the authors considered only those runs of the experiment
for which neither detector $K$ nor $L$ fires ($K=L=0$). Analogously,
in our interferometric experiment, we have discarded all those joint
detection cases for which photon 2 gets absorbed inside the phase
shifter when the configuration of the inner interferometer is
$\Phi_2$. We refer to this kind of proof as `formal' because all four
Hardy's nonlocality conditions (\ref{HM1})-(\ref{HM4}) are formally
fulfilled for the maximally entangled state. However, since this
approach involves a postselection procedure then, at least, one might
reasonably doubt that this class of experiments indeed constitutes a
valid test for nonlocality. In this Letter we have shown that, in
fact, this class of experiments does not provide us with such a
test. Indeed, if one wants to regard these experiments as true tests
of local realism, one should consider the entire pattern of joint
detection events since there is necessarily a {\em nonzero\/}
probability that photon 2 is detected by the absorber when the
configuration of the inner interferometer is $\Phi_2$, or that either
detector $K$ or $L$ does fire for any given run of the experiment
\cite{Cereceda97a}. But, as we have shown, the resulting Bell-type
inequality that obtains when the total set of localization
correlations is considered, is not violated in any case by the
quantum-mechanical predictions. It is therefore concluded that, in
spite of the fact that it is indeed possible to obtain a formal
Hardy-type nonlocality contradiction for the maximally entangled
state, the class of experiments exhibiting this kind of contradiction
cannot be used to rule out the assumption of local realism. This
situation illustrates the problem of subensemble postselection (see,
for example, Ref.\ \cite{Peres97}): whilst no local-realism violations
arise for the whole ensemble of emitted photon pairs, a Bell-type
inequality may be violated as a result of faulty (postselected)
statistics.

It is worth mentioning that a somewhat similar situation to that
examined in this Letter arises in the context of `entangled
entanglement' \cite{KZ,Cereceda97b}. In this case we have an ensemble
of three-particle systems described by the Greenberger-Horne-Zeilinger
state \cite{GHZ,GHSZ}. Spin measurements along arbitrary directions
are performed on the particles in spacelike separated regions by three
observers. Then it can be shown that the correlation function $E_{12}$
obtained by unconditionally averaging the product of the results of
the measurements on, say, particles 1 and 2 factorises into a product
$E_{12} = E_1 E_2$, and, therefore, it will be unable to yield a
violation of Bell's inequality. By unconditionally we mean that {\em
all\/} the measurement results for particles 1 and 2 are analysed,
irrespective of the result obtained in the corresponding measurement
on particle 3. Suppose, however, that one decides to analyse the
results for particles 1 and 2 {\em only\/} when observer 3 obtains the
result, say, $+1$ in the corresponding measurement on particle
3. Within this selected subensemble of measurement results for
particles 1 and 2, the resulting correlation function $E^{+}_{12}$ can
yield a violation of Bell's inequality for a suitable choice of
measurement directions. Needless to say, this procedure to get
local-realism violations rests on a biased statistical protocol and,
thereby, once again, one runs into the problem of subensemble
postselection.

We finally remark that there is still another interpretation of the
experiment at issue which does not rely on the concept of subensemble
postselection \footnote{The inspiration for this interpretation
arose out of an analysis of the experiment of Wu et al.\ made by A.\
Cabello in Section 3 of \cite{Cabello}.}\!. So, referring to Fig.\ 1,
let us suppose there is a device inside the interferometer such that,
for any emitted pair of photons emerging from the source $S$, it
prevents photon 1 from exiting the beam splitter $H_1$ whenever its
accompanying photon 2 is absorbed by the phase shifter. The remaining
pairs of photons for which photon 2 is not detected by $A_2$ are not
affected at all by the device. Thus we may think of the whole
arrangement of Fig.\ 1 (excluding the final detectors $L_i$ and $U_i$)
as a source of pairs of correlated photons in which photon 1 exits
beam splitter $H_1$ and, and the same time, photon 2 exits beam
splitter $H_2$. We will refer to this latter source as the `secondary'
source, to be distinguished from the down-conversion crystal $S$, or
`primary' source. All photons 1 (2) emerging from the secondary source
reach detectors $L_1$ or $U_1$ ($L_2$ or $U_2$). The intensity of this
source is $\frac{1}{2} (1+u^2)$ times the intensity of the primary
source. Without loss of generality, we may take the intensity of $S$
to be unity (in arbitrary units), so that the intensity of the
secondary source is simply $\frac{1}{2} (1+u^2)$. Now, the quantum
state of the two photons just before entering the final detectors can
be expressed as
\begin{equation}
|\psi \rangle = |\psi(L_1,L_2)\rangle + |\psi(L_1,U_2)\rangle
+ |\psi(U_1,L_2)\rangle + |\psi(U_1,U_2)\rangle ,  \label{state}
\end{equation}
where, for example, the term $|\psi(L_1,L_2)\rangle$ corresponds to
cases where photon 1 is detected by $L_1$ and photon 2 is detected by
$L_2$. It is important to realize that, for $u^2 <1$, the state in
Eq.\ (\ref{state}) is {\em not\/} normalised since the intensity
radiated by the secondary source is less than unity. Moreover, for the
considered case in which the conditions in Eqs.\
(\ref{HM1})-(\ref{HM4}) are fulfilled for the initial state
(\ref{ME}), it can be shown that the unnormalised state (\ref{state})
does {\em not\/} produce any violation of the CHSH inequality,
\begin{equation}
\left| E(\Phi_1,\Phi_2) + E(\Phi_1,\Phi^{\prime}_2) +
E(\Phi^{\prime}_1,\Phi_2) - E(\Phi^{\prime}_1,\Phi^{\prime}_2) \right|
\leq 2.  \label{CHSH}
\end{equation}
It is only if normalisation of the state (\ref{state}) is imposed that
the wave function of the pair of photons emerging from the secondary
source will lead to a violation of inequality (\ref{CHSH}). As a
matter of fact, for the experimental set-up considered in Fig.\ 1, the
normalisation of the state (\ref{state}) amounts to `erasing' all the
events for which photon 2 is detected by the absorber, so that, in a
sense, only those events for which both photons of a pair impinge on
the final detectors have physical reality (indeed, if we think of the
secondary source as a kind of black box, this will be the impression
experienced by any observer standing outside the box, since all what
such an observer `sees' are photons emanating {\em in pairs\/} from
the box). This interpretation is consistent with the analysis made in
Section 4 where we saw that, if one just cuts out the last term on the
right-hand side of the inequality (\ref{CH-total}), then this latter
inequality is automatically violated by the relevant predictions
(\ref{HM1})-(\ref{HM4}) that quantum mechanics makes for the maximally
entangled state (\ref{ME}).  \enlargethispage*{.2cm}

The interpretation of the experiment presented in the last paragraph,
however, is not equivalent to that explained in the rest of the Letter
(see, in particular, the last two paragraphs of Section 3 and Section
4). This is because, for the considered case in which the conditions
(\ref{HM1})-(\ref{HM4}) are satisfied for the state (\ref{ME}), the
quantum-mechanical violation of the inequality (\ref{CH-total}) rests
on subensemble postselection whereas the violation associated with the
inequality (\ref{CHSH}) essentially involves a preselection procedure,
in that the selection of the CHSH-violating subensemble out of the
original ensemble takes place {\em before\/} proceeding to the final
measurements. Furthermore, the CH-type inequality in Eq.\ (\ref{CH})
can be violated by unnormalised probabilities $P(L_{1},\Phi
_{1};L_{2},\Phi _{2})$, $P(U_{1},\Phi _{1};U_{2},\Phi _{2}^{\prime
})$, $P(U_{1},\Phi _{1}^{\prime };U_{2},\Phi _{2})$, and $P(U_{1},\Phi
_{1}^{\prime };U_{2},\Phi _{2}^{\prime })$, because, in contrast with
the CHSH inequality (\ref{CHSH}), it involves only {\em ratios\/} of
probabilities, rather than their absolute magnitudes
\cite{C-S}.\footnote{This is most easily seen when we write the
inequality (\ref{CH}) as, $P(U_{1},\Phi _{1}^{\prime };U_{2},\Phi
_{2}^{\prime }) - P(L_{1},\Phi _{1};L_{2},\Phi _{2}) - P(U_{1},\Phi
_{1};U_{2},\Phi _{2}^{\prime }) - P(U_{1},\Phi _{1}^{\prime
};U_{2},\Phi _{2}) \leq 0$. Clearly, as this inequality has a zero
value as its upper bound, it is insensitive to the overall
normalisation of probabilities.}  In any case, both approaches
together show up the fact that, if one wants to obtain a Hardy-type
contradiction for the maximally entangled state, then one must either
sift the measurement data according to some given procedure or else
manipulate the original two-photon state emerging from the primary
source before the photons arrive at the final detectors.
\vspace{-.2cm}

\ack{The author wishes to thank Ad\'{a}n Cabello for useful comments
and discussions. He would also like to thank an anonymous referee for
making a number of valuable suggestions which led to an improvement of
an earlier version of this Letter.}


\newpage

\begin{thebibliography}{99}
\bibitem{Hardy93} L. Hardy, Phys. Rev. Lett. 71 (1993) 1665.
\bibitem{Goldstein94} S. Goldstein, Phys. Rev. Lett. 72 (1994) 1951.
\bibitem{CB} A. Chefles and S.M. Barnett, Phys. Lett. A 236 (1997) 177.
\bibitem{Cereceda98} J.L. Cereceda, Found. Phys. Lett. 12 (1999) 211. This paper is also available at Los Alamos e-print archive, quant-ph/9908039.
\bibitem{Wu et al.} X. Wu, R. Xie, X. Huang and Y. Hsia, Phys. Rev. A 53 (1996) R1927.
\bibitem{Cereceda97a} J.L. Cereceda, Phys. Rev. A 55 (1997) 3968.
\bibitem{Yurke-Stoler} B. Yurke and D. Stoler, Phys. Rev. A 46 (1992) 2229.
\bibitem{Popescu+Gisin} S. Popescu, Phys. Rev. Lett. 74 (1995) 2619; N. Gisin, Phys. Lett. A 151 (1996) 210.
\bibitem{Peres96} A. Peres, Phys. Rev. A 54 (1996) 2685.
\bibitem{Mermin} N.D. Mermin, Am. J. Phys. 62 (1994) 880. See also a contribution of this author, in: Fundamental Problems in Quantum Theory: A Conference held in Honor of Professor John A. Wheeler, eds. D.M. Greenberger and A. Zeilinger, Ann. N.Y. Acad. Sci. 755 (1995) 616.
\bibitem{Hardy94} L. Hardy, Phys. Rev. Lett. 73 (1994) 2279.
\bibitem{C-H} J.F. Clauser and M.A. Horne, Phys. Rev. D 10 (1974) 526.
\bibitem{C-S} J.F. Clauser and A. Shimony, Rep. Prog. Phys. 41 (1978) 1881.
\bibitem{PHZ} S. Popescu, L. Hardy and M. \.{Z}ukowski, Phys. Rev. A 56 (1997) R4353.
\bibitem{Santos} E. Santos, Phys. Rev. Lett. 66 (1991) 1388; Phys. Rev. A 46 (1992) 3646.
\bibitem{Caro-G} L. De Caro and A. Garuccio, Phys. Rev. A 50 (1994) R2803.
\bibitem{Peres97} A. Peres, in: Potentiality, Entanglement and
Passion-at-a-Distance: Quantum Mechanical Studies for Abner Shimony,
vol. 2, eds. R.S. Cohen, M.A.~Horne and J. Stachel, Kluwer Academic,
Dordrecht, The Netherlands, 1997, p. 191. This paper can also be found
in Los Alamos e-print archive, quant-ph/9512003.
\bibitem{Horne-Zeilinger} M.A. Horne and A. Zeilinger, in:
Proc. Symp. on the Foundations of Modern Physics, eds. P. Lahti and
P. Mittelstaedt, World Scientific, Singapore, 1985, p.~435. See also,
M.A. Horne, A. Shimony and A. Zeilinger, Phys. Rev. Lett. 62 (1989)
2209.
\bibitem{CHSH}J.F. Clauser, M.A. Horne, A. Shimony and R.A. Holt,
Phys. Rev. Lett. 23 (1969) 880.
\bibitem{Bell} J.S. Bell, Speakable and Unspeakable in Quantum
Mechanics, Cambridge University Press, Cambridge, 1987.
\bibitem{Jarrett} J.P. Jarrett, No\^{u}s 18 (1984) 569;
L.E. Ballentine and J.P. Jarrett, Am. J. Phys. 55 (1987) 696.
\bibitem{Shimony} A. Shimony, in: Proc. 1st Int. Symp. on Found. of
Quantum Mechanics in the Light of New Technology, eds. S. Kamefuchi et
al., Physical Society of Japan, Tokyo, 1984, p. 225. Reprinted in
A. Shimony, Search for a Naturalistic World View, vol. II, Cambridge
University Press, Cambridge, 1993, p. 130.
\bibitem{EV} A. Elitzur and L. Vaidman, Found. Phys. 23 (1993) 987.
\bibitem{Vaidman} L. Vaidman, Quantum Opt. 6 (1994) 119; Los Alamos
e-print archive, quant-ph/9610033.
\bibitem{Noh-Hong} T.G. Noh and C.K. Hong, Quantum Semiclass. Opt. 10
(1998) 637.
\bibitem{KZ} G. Krenn and A. Zeilinger, Phys. Rev. A 54 (1996) 1793.
\bibitem{Cereceda97b} J.L. Cereceda, Phys. Rev. A 56 (1997) 1733.
\bibitem{GHZ} D.M. Greenberger, M.A. Horne and A. Zeilinger, in:
Bell's Theorem, Quantum Theory, and Conceptions of the Universe,
ed. M. Kafatos, Kluwer Academic, Dordrecht, The Netherlands, 1989,
p. 69.
\bibitem{GHSZ} D.M. Greenberger, M.A. Horne, A.~Shimony and
A. Zeilinger, Am. J. Phys. 58 (1990) 1131.
\bibitem{Cabello} A. Cabello, Nonlocality without inequalities has not been proved for maximally entangled states, submitted preprint, 1999.

\end{thebibliography}
\end{document}